\documentclass[final,1p,times]{elsarticle} 
\usepackage{graphicx} 
\usepackage{amssymb} 
\usepackage{amsthm} 
\usepackage{lineno} 

\usepackage{subfigure}
\newcommand{\pt}{p_\mathrm{T}}
\newcommand{\kt}{k_\mathrm{T}}
\newcommand{\et}{E_\mathrm{T}}

\journal{Nuclear Physics A} 
\begin{document} 

\begin{frontmatter} 


\title{Initial state nuclear effects for jet production measured in $\sqrt{s_\mathrm{NN}} = 200~\mathrm{GeV}$ d+Au collisions by STAR}

\author{Jan Kapit\'an for the STAR Collaboration}

\address{Nuclear Physics Institute ASCR, Na Truhlarce 39/64, 18086 Praha 8, Czech Republic}

\begin{abstract} 
Full jet reconstruction in heavy-ion collisions is a promising tool for quantitative study of properties of the dense medium produced at RHIC. Measurements of d+Au collisions are important to disentangle initial state nuclear effects from medium-induced $k_\mathrm{T}$ broadening and jet quenching. We report measurements of mid-rapidity ($|\eta_{jet}| < 0.4|$) di-jet correlations in d+Au using high-statistics run 8 RHIC data at $\sqrt{s_\mathrm{NN}} = 200~\mathrm{GeV}$.
\end{abstract} 

\end{frontmatter} 




\section{Jet reconstruction}
\label{jets}

Recently reported results on full jet reconstruction in heavy-ion collisions~\cite{HP,QM} provide new insights into parton energy loss in hot QCD matter. Baseline measurements in smaller collision systems are needed to better interpret these results. In particular, measurement of the nuclear $\kt$ broadening~\cite{vitev}, the transverse momentum of a jet pair, in d+Au collisions is important.

The present analysis is based on $\sqrt{s_\mathrm{NN}} = 200~\mathrm{GeV}$ d+Au run 8 data from the STAR experiment. The Beam Beam Counter detector in the Au nucleus fragmentation region is used to select the 20\% highest multiplicity events.
In addition to the minimally biased trigger (MB) two high tower triggers (HT) were used, requiring one tower in the Barrel Electromagnetic Calorimeter (BEMC) over threshold: $\et > 4.3~\mathrm{GeV}$ (HT2 trigger) and $\et > 8.4~\mathrm{GeV}$ (HT4 trigger).

The BEMC is used to measure the neutral component of jets, and the Time Projection Chamber (TPC) detector is used to measure the charged particle component of jets. An upper $\pt < 15~\mathrm{GeV/}c$ cut is applied to TPC tracks due to uncertainties in TPC tracking performance at high-$\pt$. 
In the case of a TPC track pointing to a BEMC tower, its momentum is subtracted from the tower energy to avoid double counting (electrons, MIP and possible hadron showers in the BEMC). 
The tracks and towers are selected in pseudo-rapidity $|\eta| < 0.9$. To reduce underlying event background, a cut $\pt > 0.5~\mathrm{GeV/}c$ was used for tracks and towers. To reduce possible BEMC backgrounds, the jet neutral energy fraction is required to be within $(0.1,0.9)$.

Recombination jet algorithms kt and anti-kt from the Fastjet package~\cite{fj} are used with a resolution parameter $R = 0.5$. This sets the jet fiducial acceptance to $|\eta| < 0.4$. 
To subtract the contribution of d+Au underlying event background, a method based on active jet areas is used~\cite{bgsub}. Background density, obtained using Fastjet, is $\approx 1~\mathrm{GeV}/c$ per unit area.  
Due to the asymmetry of the colliding d+Au system, the background is asymmetric in $\eta$. This dependence was fit with a linear function of $\eta$, and is included in the background subtraction procedure. 

To study detector effects, Pythia 6.410 and GEANT detector response simulations were used. Jet reconstruction was performed at MC level (PyMC) and at detector level (PyGe). To study effects of the d+Au background, PyGe events were added to MB-triggered d+Au events (PyBg).

\begin{figure}[htb]
\begin{minipage}[h]{0.29\textwidth}  
\centering
\includegraphics[width=\textwidth]{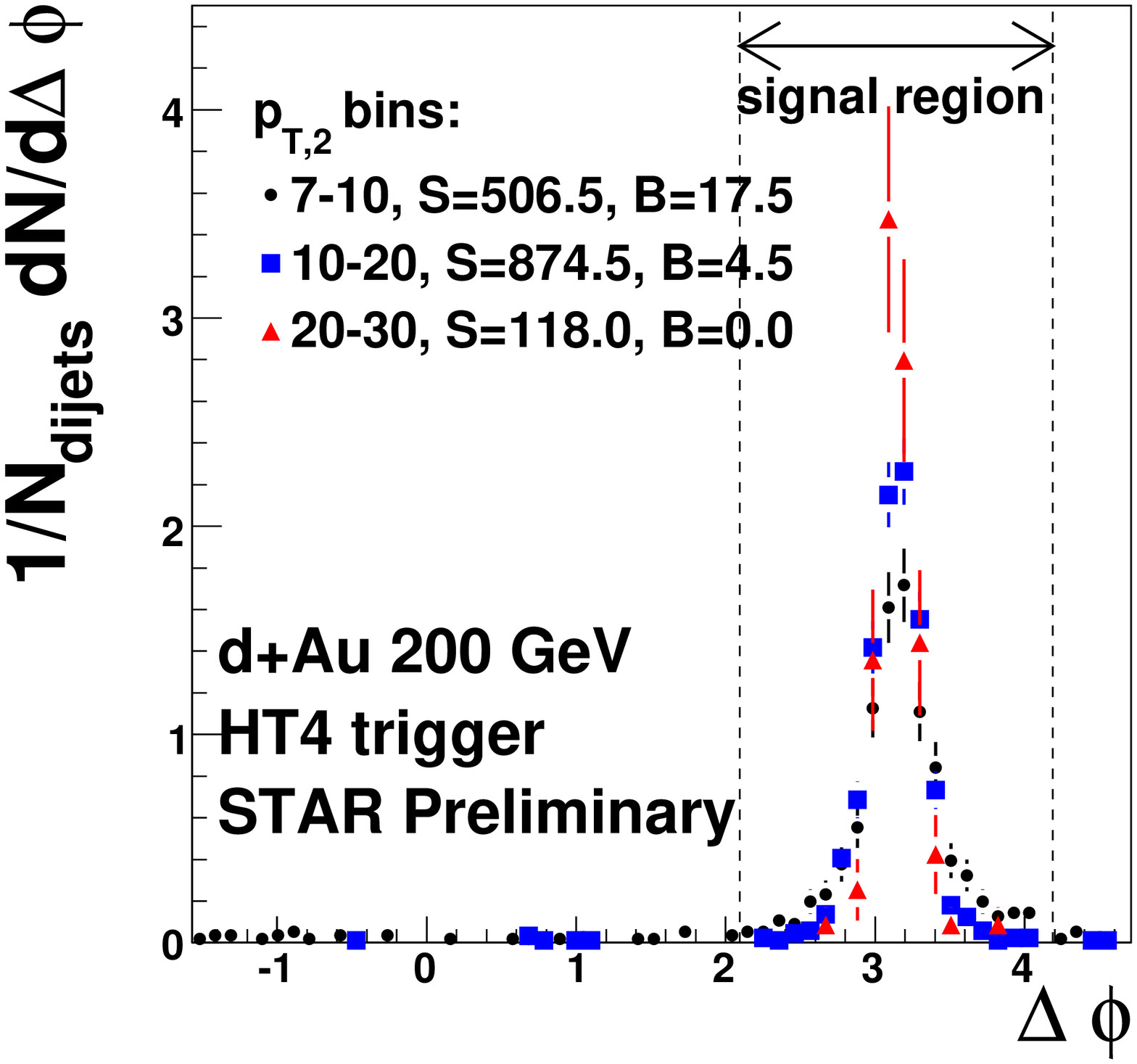}
\vspace{-24pt}
\caption{Di-jet $\Delta\phi$ distribution for anti-kt algorithm.}
\label{fig:dphi}
\end{minipage}
\hfill
\begin{minipage}[h]{0.67\textwidth}
\includegraphics[width=\textwidth]{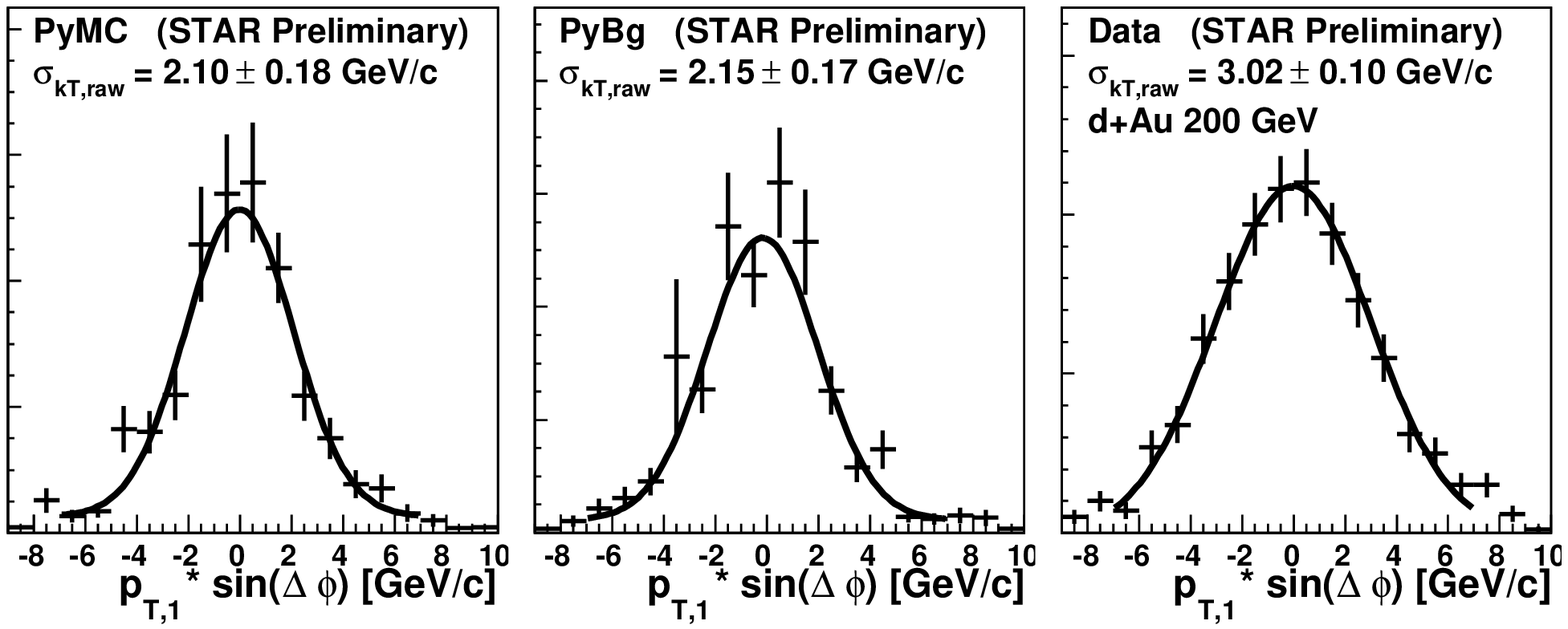}
\vspace{-23pt}
\caption{Distributions of $k_\mathrm{T,raw} = p_\mathrm{T,1} \sin(\Delta\phi)$ for anti-kt algorithm and HT4 trigger, $10 < p_\mathrm{T,2} < 20~\mathrm{GeV/}c$.}
\label{fig:kt}
\end{minipage}
\hfill
\end{figure}

\section{Di-jet results}
\label{results}
The two highest energy jets ($p_\mathrm{T,1} > p_\mathrm{T,2}$) in each event were used for di-jet analysis. Figure~\ref{fig:dphi} shows the di-jet $\Delta\phi$ distribution (B: total number of background di-jets). Clear di-jet correlations are observed and the background is decreasing with rising $p_\mathrm{T,2}$. Acceptance effects on di-jet $\Delta\phi$ were studied by randomly sampling the jet $\phi$ distribution, and are negligible ($\approx 1\%$) for $\Delta\phi \approx \pi$.

Distributions of $k_\mathrm{T,raw} = p_\mathrm{T,1} \sin(\Delta\phi)$ for PyMC, PyBg and Data were constructed for di-jets. Gaussian $\sigma_{k_\mathrm{T,raw}}$ was measured for the two jet algorithms, HT2 and HT4 triggers and two ($10 - 20~\mathrm{GeV/}c$, $20 - 30~\mathrm{GeV/}c$) $p_\mathrm{T,2}$ bins. Figure~\ref{fig:kt} shows an example of these distributions. Interestingly, the sigma widths of PyMc and PyBg $k_\mathrm{T,raw}$ distributions are similar, due to the interplay between jet $\pt$ and $\Delta\phi$ resolutions. The measured $\sigma_{k_\mathrm{T,raw}} = 3.0 \pm 0.1~\mathrm{(stat)}~\mathrm{GeV/}c$. 

The systematic errors coming from a small dependence on the $|\Delta\phi - \pi|$ cut for back-to-back di-jet selection (varied between 0.5 and 1.0) and differences between the various selections (trigger, $p_\mathrm{T,2}$ range, jet algorithm) are estimated to be $0.2~\mathrm{GeV/}c$.
Additional systematic effects for this measurement include the TPC performance in the high luminosity environment in run 8 (currently under study) and the fact that the final BEMC calibration is not yet finished. We estimate systematic uncertainty from the latter two to be of order $10\%$ ($0.3~\mathrm{GeV/}c$). Total systematic uncertainty is thus $\sqrt{0.2^{2} + 0.3^{2}}~\mathrm{GeV}/c \approx 0.4~\mathrm{GeV}/c$. 

\section{Summary}
\label{summary}
Di-jet $\kt$ width in highest multiplicity d+Au collisions was measured to be $\sigma_{k_\mathrm{T,raw}} = 3.0 \pm 0.1 \mathrm{(stat)} \pm 0.4 \mathrm{(syst)}~\mathrm{GeV/}c$. Di-jet studies in different centrality d+Au collisions and p+p collisions are needed to assess possible nuclear contributions to measured $\kt$.



\section*{Acknowledgement}
\label{acknowledgement}

This work was supported in part by GACR grant 202/07/0079 
and by grants LC07048 and LA09013 of the Ministry 
of Education of the Czech Republic.

\end{document}